\documentclass[11pt]{article}
\pdfoutput=1  
\usepackage{graphicx}

\usepackage{appendix}

\usepackage{amssymb}
\usepackage{amsmath}
\usepackage{amsfonts}
\usepackage{upgreek}
\usepackage{latexsym}
\usepackage{slashed,upgreek,amscd,cancel,tensor}
\usepackage{adjustbox}
\usepackage{epsfig}
\usepackage{cite}
\usepackage{hyperref}
\usepackage{amsfonts}
\usepackage{booktabs}
\usepackage{siunitx}
\usepackage{caption}
\usepackage{float}
\usepackage{color}

\numberwithin{equation}{section}

\definecolor{MyBlue}{rgb}{0.15,0.15,0.70}

\hypersetup{
colorlinks=true,
citecolor=MyBlue,
linkcolor=MyBlue,
urlcolor=MyBlue
}

\input epsf
\usepackage[top=2cm,bottom=3cm,left=2.6cm,right=2.6cm,asymmetric]{geometry}

\newcommand{\be}{\begin{equation}}
\newcommand{\ee}{\end{equation}}
\newcommand{\beq}{\begin{equation}}
\newcommand{\eeq}{\end{equation}}
\newcommand{\bea}{\begin{eqnarray}}
\newcommand{\eea}{\end{eqnarray}}

\usepackage[stable]{footmisc}

\def\dkmu2{\delta K_{\mu \nu}\delta K^{\mu \nu}}
\def\pmu2{  \phi_{\mu \nu}\phi^{\mu \nu}}

\renewcommand\[{\left[}

\newcommand\ees{\end{eqnarray}}
\newcommand\bees{\begin{eqnarray}}

\newcommand\h{h}

\begin{document}

\begin{center}
\LARGE{\bf Deformed General Relativity \\ and Quantum Black Holes Interior}
\\[1cm] 

\large{Denis Arruga$^{\rm a}$,  Jibril Ben Achour$^{\rm b}$, Karim Noui$^{{\rm c},{\rm a}}$}
\\[0.5cm]

\small{
\textit{$^{\rm a}$
  APC -- Astroparticule et Cosmologie, 
\\Universit\'e Paris Diderot Paris 7, 75013 Paris, France}}
\vspace{.2cm}

\small{
\textit{$^{\rm b}$ Center for Gravitational Physics, Yukawa Institute for Theoretical Physics \\ [0.05cm]
Kyoto University, Kyoto, Japan}}

\vspace{.2cm}

\small{
\textit{$^{\rm c}$ Institut Denis Poisson, \\ [0.05cm]
Universit\'e de Tours, Universit\'e d'Orl\'eans,  
\\[0.05cm]
Parc de Grandmont, 37200 Tours, France}}

\end{center}

\vspace{.2cm}

\begin{abstract}
Effective models of black holes interior have led to several proposals for regular black holes. In the so-called polymer models,  based on effective deformations of the phase space of spherically symmetric general relativity in vacuum, one considers a deformed Hamiltonian constraint while keeping a non-deformed vectorial constraint, leading under some conditions to a notion of deformed covariance. In this article, we revisit and study further the question of covariance in these deformed gravity models. In particular, we propose a Lagrangian formulation for these deformed gravity models where polymer-like deformations are introduced at the level of the full theory prior to the symmetry reduction and prior to the Legendre transformation. This enables us to test whether the concept of deformed covariance found in spherically symmetric vacuum gravity can be extended to the full theory, and we show that,  in the large class of models we are considering, the deformed covariance can not be realized beyond spherical symmetry in the sense that the only deformed theory which leads to a closed constraints algebra is general relativity. Hence, we focus  on the spherically symmetric sector, where there exist non-trivial deformed but closed constraints algebras. We investigate the possibility to deform the vectorial constraint as well and we prove that 
non-trivial deformations of the vectorial constraint with the condition that the constraints algebra remains closed do not exist. 
Then, we compute the most general deformed Hamiltonian constraint which admits a closed constraints algebra and thus leads to a well-defined effective theory associated with a notion of deformed covariance. Finally, we study static solutions of these effective theories and, remarkably, we solve explicitly and in full generality the corresponding modified Einstein equations, even for the effective theories which do not satisfy the
 closeness condition. In particular, we give the expressions of the components of the effective metric (for spherically symmetric black holes interior) in terms of the functions that govern the deformations of the theory.

\end{abstract}
\vspace{.2cm}
\newpage
\tableofcontents
\newpage

\section{Introduction}

Black hole are iconic predictions of general relativity. Their classical description is very well established, and strong evidences for their existence is now regularly reported through the gravitational wave astronomy. Yet, their quantum description is still an open problem. The presence of a classical singularity in their core prevents from having a complete description of their interior geometry, and more generally of their formation. It is widely believed that, once a consistent UV completion of general relativity will be available, classical singularities will be replaced by a well defined  ``quantum geometry'', allowing for a consistent unitary evolution of the degrees of freedom at the quantum level. Since a complete and consistent quantum theory of gravity is still missing, important efforts towards the understanding of the quantum description of black hole have been focused on developing instead effective approaches. The construction of effective regular black holes geometries have triggered an important activity over the last decades, initiated by the work of Bardeen (see \cite{Carballo-Rubio:2018jzw} for a review). However, such ad hoc models suffer from several inconsistencies, and in particular, fail to provide a framework to discuss completely the end point of the evaporation\footnote{See however \cite{Simpson:2019mud} for a new construction with potentially interesting features.}, as they predict generically an infinite time of evaporation \cite{Carballo-Rubio:2018pmi}.

Alternative approaches to quantum black holes rely on canonical quantizations of the phase space of spherically symmetric  gravity in vacuum (see \cite{Kuchar:1994zk, Thiemann:1992jj,Thiemann:1993ym} for early works on this topic).  The polymer quantization, inspired by loop quantum gravity, has led to several proposals for the effective description of the Schwarzschild interior. Early works in this direction, motivated by previous results on the polymer quantization of cosmological backgrounds, were presented in \cite{Bojowald:2004af, Bojowald:2004ag, Bojowald:2005cb, Ashtekar:2005qt, Modesto:2006mx, Bohmer:2007wi, Boehmer:2008fz, Modesto:2009ve, Chiou:2012pg}, suggesting new scenarii for the end of the collapse \cite{Ashtekar:2005cj, Bojowald:2005qw}. Since then, the polymer quantization of this model has been revisited by several authors \cite{Gambini:2013ooa, Gambini:2013hna, Gambini:2014qta, Brahma:2014gca, Campiglia:2016fzp}, with several improvements along the years. Later on, effective approaches based on a ``quantum'' deformation of the classical phase space were used to provide effective and regular metrics  for the Schwarzschild interior. Such effective models appear as a shortcut to investigate possible quantum corrections to the interior geometry without going through the whole quantization procedure\footnote{The common underlying assumption of these effective polymer models is that the modified hamiltonian constraint actually corresponds to the expectation value of the hamiltonian operator in some suitable coherent states at the quantum level. However, it is fair to say that the concrete realization of this assumption for polymer black holes  is usually left aside, and one only deals with a modified classical system. Recently, efforts towards computing this expectation value of the hamiltonian operator for spherically symmetric quantum geometries were presented in \cite{Alesci:2018loi}, based on the framework of quantum reduced loop quantum gravity. Consequences for the black hole interior geometry were discussed in \cite{Alesci:2019pbs}. }. Following this strategy, one found effective geometries where the singular vacuum Schwarzschild interior has been replaced by a black-to-white hole bounce, first in \cite{Corichi:2015xia, Olmedo:2017lvt}, and more recently in \cite{Ashtekar:2018lag,  Ashtekar:2018cay}. 
These new solutions have been discussed in \cite{Bodendorfer:2019xbp, Bouhmadi-Lopez:2019hpp, Bojowald:2019dry} and 
an alternative effective polymer model of black-to-white hole transition, providing several improvements, has been introduced in \cite{Bodendorfer:2019cyv, Bodendorfer:2019nvy, Bodendorfer:2019jay} (see also \cite{Ziprick:2016ogy, Cortez:2017alh, Lobo:2018fym, Protter:2018tbj, Vakili:2018xws, Amirfakhrian:2018ehm, Moulin:2018uap, Morales-Tecotl:2018ugi, Barrau:2018rts, Assanioussi:2019twp} for additional investigations on polymer black hole geometries and \cite{Rovelli:2014cta, Haggard:2014rza, Bianchi:2018mml, Barcelo:2015uff} for yet another class of effective models of black-to-white hole transition). While these classical polymer constructions provide  a straightforward and interesting platform to investigate quantum corrected effective metrics, the strategy employed in these hamiltonian constructions turns out to suffer from several shortcomings related to the covariance of the quantum corrections.

To explain this problem, we start giving the basic building bocks  of these polymer effective models. First, let us recall that the hamiltonian formulation of vacuum spherically symmetric gravity, where there are no local degrees of freedom, leads to two first class constraints, the hamiltonian and (radial) vectorial constraints, 
which generate time and radial diffeomorphisms. Then, the common strategy employed in these effective polymer models is to introduce some regularization at the phase space level through some modifications of the hamiltonian constraint known as point-wise holonomy corrections, while keeping the vectorial constraint unchanged. In general, this regularization breaks the covariance of the system, as the algebra of the first class constraints is no longer closed. Then the solutions to the modified equations of motion one can extract from this modified hamiltonian system do not have a classical space-time interpretation, as covariance is lost. In models such as \cite{Corichi:2015xia, Olmedo:2017lvt, Ashtekar:2018lag,  Ashtekar:2018cay} as well as in \cite{Bodendorfer:2019cyv}, where only the homogeneous interior region is considered, the issue of covariance is not visible.  However, if one tries to extend these models to take
into account inhomogeneities, the issue of the covariance breaking becomes important.

Interestingly, it was realized in a series of articles \cite{Bojowald:2008ja, Bojowald:2008bt, Bojowald:2009ih, Bojowald:2011js, Bojowald:2011aa, Bojowald:2012ux} that under suitable conditions, one can construct regularizations where the deformed algebra of constraints remain closed\footnote{Even more recently, it was realized that if one starts from the spherically symmetric vacuum gravity phase space in term of the self-dual variables, for which the Barbero-Immirzi parameter $\gamma =\pm i$, one obtains polymer-like modifications while keeping the algebra of constraints closed and undeformed \cite{BenAchour:2016brs}. This construction was generalized to the Gowdy model and perturbative inhomogeneous cosmological background. Moreover, the polymer-like modifications can be implemented within the $\bar{\mu}$-scheme \cite{BenAchour:2017jof, BenAchour:2016leo, Wu:2018mhg}. These results have been generalized recently in \cite{Bojowald:2019fkv}. }. The consequences of this deformed covariance was then investigated in depth in the context of spherically symmetric backgrounds \cite{Tibrewala:2013kba, Tibrewala:2012sy, Bojowald:2015zha} and then extended to more general symmetry reduced models in \cite{Cailleteau:2011kr, Barrau:2014maa, Bojowald:2015sta, Bojowald:2016hgh, Bojowald:2016vlj, Bojowald:2016itl}. Phenomenological implications were explored both for inhomogeneous cosmological backgrounds as well as for black holes in \cite{Bojowald:2015gra, Barrau:2016sqp, Bojowald:2014zla}. More recently, exact solutions for black holes interior in a very large class of gravitational systems with deformed covariance were found in \cite{BenAchour:2018khr, Bojowald:2018xxu}, and relations between these deformed
theories and modified gravity (in the framework of scalar-tensor theories) have been investigated in \cite{BenAchour:2017ivq, Ghersi:2017ley, Carballo-Rubio:2018czn, Cuttell:2018axb, Cuttell:2018gai, Cuttell:2019fgl, Cuttell:2019rbm}.

\medskip

In this paper, we further investigate the issue of covariance in effective black holes models.  As we have already said, there exist modifications
of spherically symmetric general relativity, inspired from loop quantum gravity, which lead to a deformed but closed constraints algebra. This property makes the effective theories particularly interesting because they are invariant under a  deformed covariance, 
thus there is no effective quantum anomalies, and then the classical solutions have a clear geometrical meaning. 
Now, we can ask the question whether one can extend such a construction beyond spherical symmetry.
We explore this question in section~\ref{sectionII}. 
We consider a large class of modified theories of gravity whose deformation is not only inspired from (holonomy corrections induced by) 
loop quantum gravity but is also a natural generalization of the deformation introduced in the context of spherical symmetry. In particular, 
these theories are invariant under three-dimensional space-like diffeomorphisms. We prove that any theory of this class which has a closed
constraints algebra reduces, after some canonical transformations, to general relativity with a cosmological constant. Hence, at the level of the full theory, we show that 
there is no non-trivial deformations of general relativity in a large class of modified gravity theories with the condition that they produce 
a closed constraints algebra. On one hand, this
result clearly contrasts with what happens when space-times are reduced to spherical symmetry\footnote{Or cylindrical symmetry.}. On the other hand, this finding parallels the well known Hojman-Kuchar-Teteilboim theorem which allows to reconstruct uniquely, from the Dirac's hypersurface algebra, the Einstein-Hilbert action \cite{Hojman:1976vp}. 

In section \ref{sectionIII}, we revisit and study new aspects of  loop (holonomy) deformations of spherically symmetric general relativity.
We start with a quick Hamiltonian analysis of general relativity reduced to spherical symmetry using Ashtekar like variables, and
we obtain the two expected constraints, the Hamiltonian constraint and the radial diffeomorphisms constraint. Then, we ask the question
whether one can deform these constraints to keep their algebra closed. First, we show that there is no non-trivial deformations 
(up to canonical transformations) of the sub-algebra generated by the radial diffeomorphisms constraint. In other words, space-like 
diffeomorphisms cannot be consistently deformed in an effective theory of general relativity and they remain, in that sense, classical. This
result is consistent with loop quantum gravity where kinematical quantum states are invariant under classical space-like diffeomorphisms. 
However, it is possible to deform the Hamiltonian constraint in a way that the constraints algebra is deformed but closed. We compute
and recover the most general deformation which keeps the algebra closed. 

In section \ref{sectionIV}, we compute the equations of motion for the effective metric components and we look for their  solutions 
corresponding to ``static'' black holes interiors, which means that the metric components are time dependent only. Notice that,  in that case, 
the closeness of the constraints algebra is not an issue and one can safely relax this condition. Hence, we consider in this section 
very general deformations of the Hamiltonian constraint with the only restriction that it still transforms as a scalar under radial diffeomorphisms. 
Remarkably, we completely solve the deformed Einstein equations in that very general case and we give the expressions of the components
of the effective metric in terms of the functions that govern the deformations. For concreteness, we illustrate this general result with 
simple examples. 

We conclude the article with a brief summary and a discussion of the results. We also present some interesting perspectives.

  \section{Deformed General Relativity}

\label{sectionII}

In this section, we would like to recall some aspects and also give new results on modified gravity which will be useful in 
the following sections when we study modifications of gravity induced by loop quantum gravity. In particular, we start by 
introducing a large class of theories of modified gravity which share some features with  effective theories of loop quantum gravity
when it is reduced to spherical symmetry. However, contrary to effective loop quantum gravity,  these modified theories of gravity are
defined in full generality and not only in the case of  symmetry reduced backgrounds. Then, we discuss these theories and show 
in particular the impossibility to have a closed deformed algebra of constraints in 3+1 dimensions in this class of theories. 
This result raises the question
of the possibility to extend the construction of effective loop quantum gravity  theories with a closed deformed algebra of constraints 
beyond spherical symmetry. 

\subsection{From (effective) loop quantum gravity to modified gravity}
As any theory of quantum gravity, loop quantum gravity is expected to modify Einstein equations at the effective level. Such modifications have
been computed and studied for homogeneous and spherically symmetric space-times to understand the effects of quantum gravity in early cosmology and black holes physics. In general, one distinguishes between two types of deformations, known as 
holonomy and inverse-triad corrections. Here we are concerned only with (point-) holonomy corrections which are  local modifications in the 
sense that they do not produce non-local equations of motion. 

These (point-) holonomy corrections are assumed to modify only the Hamiltonian constraint and do not affect the vectorial constraints.  
Indeed, in the construction of the kinematical Hilbert space of loop quantum gravity, the invariance under diffeomorphisms is imposed
``classically'' from the action of diffeomorphisms on the graphs of spin-network states.  Hence, in general, one does not use holonomies 
to quantize the vectorial constraints.  For this reason, it is natural to require that any effective theory of loop quantum gravity remains 
invariant  under (undeformed) spatial diffeormorphims.  Furthermore, we will show in the case of spherical symmetry that it is impossible
to construct a non-trivial deformation of the vectorial constraint which leaves the deformed diff-algebra closed (see section \ref{DefofDiff}).

To construct such effective theories, we assume that the space-time $\cal M$ if of the form $\Sigma \times \mathbb R$ where $\Sigma$ is
a space-like slice, and we consider the ADM parametrization of the metric,
\bea
ds^2&=&-N^2 dt^2 +\gamma_{ij} (dx^i+N^i dt)(dx^j+N^j dt) \, ,
\label{metric_ADM}
\eea
where $\gamma_{ij}$ is the spatial metric on $\Sigma$, $N$  is the lapse function and $N^i$ the shift vector. 
We also need to introduce the second fundamental form (extrinsic curvature tensor) whose components are  given by 
\beq
K_{ij}=\frac{1}{2N}\left( \dot{\gamma}_{ij}-D_iN_{j}-D_jN_{i}\right)\,,
\label{Eij}
\eeq
where  $D_i$ denotes the spatial covariant derivative associated with 
the spatial metric $\h_{ij}$. In this parametrization, the Einstein-Hilbert action takes the form
\bea
\label{EHaction}
S_{EH}[g] = \int d^4x \, \sqrt{- g } \, {\cal R} \; = \; \int d^4x \; N \, \sqrt{\gamma} \left( K_{ij} K^{ij} - K^2 + R \right) \, ,
\eea
where $\cal R$ and $R$ respectively the 4-dimensional and the 3-dimensional Ricci scalars, $g$ and $\gamma$ are the 
determinant of the metrics $g_{ij}$ and $\gamma_{ij}$ respectively. {In the following, we shall omit the boundary contribution to the action even though a complete analysis would require to include them.}

To mimic  loop quantum gravity type 
modifications, we now introduce a class of modified theories of gravity defined by an action of the form
\bea
\label{LQGmod}
S[\gamma_{ij},N,N^i] \equiv \int d^4x \; N \sqrt{\gamma} \left[ F(K^i_j) + R \right]\, ,
\eea
where $F$ is an arbitrary 3-dimensional scalar constructed from the components of the extrinsic curvature $K_{i}^j$. Only the part 
of the action which depends on $K_i^j$ is modified. This choice is motivated by the fact that, as we are going to recall later on (for spherically symmetric spacetimes), (point-) holonomy corrections affect mainly the components of the extrinsic curvature, whereas the components of the
3-dimensional Ricci tensor are left unchanged\footnote{Notice that the standard boundary term has been ignored here, but it would be interesting to investigate its role in such deformed theory. A careful investigation of this boundary term is nevertheless needed to properly understand how the quasi-local observables are affected in such deformed gravity theory.}. Interestingly, these modifications contrast with deformations \`a la Horava gravity \cite{Horava:2009uw, Blas:2009yd} where 
the spatial ``Ricci" part of the action is strongly modified whereas the ``K-part" is (almost) unchanged compared to general relativity (up to a detuning relative coefficient between $K_{ij} K^{ij}$ and $K^2$). 

Hence, we consider the actions \eqref{LQGmod} as models for effective loop quantum gravity theories in any background. As $K_j^i$ can be viewed as a 3-dimensional matrix, the most general function $F(K^i_j)$ which transforms as a scalar under space-like diffeomorphisms can be written as
\bea
\label{funF}
F(K^i_j) \; = \; F\left(\text{tr}(K),\text{tr}(K^2),\text{tr}(K^3)\right) \, ,
\eea
where $\text{tr}(M) \equiv M_{ij} \gamma^{ij} = M_i^i$ for any matrix $M$. This is a direct consequence of the Cayley-Hamilton theorem. For simplicity, we are using the same notation for the two (different) functions F in the l.h.d. and r.h.s. of \eqref{funF}.

Before analyzing further this class of  theories, let us make a couple of remarks. First, modified theories of gravity which are invariant under spatial diffeomorphisms only have been studied intensively these last years on many aspects.  Their Lagrangians involve not only the extrinsic curvature tensor, but also the 3-dimensional Ricci tensor, the lapse and the shift, and their covariant derivatives. Horava gravity is an example of such theories with remarkable ultra-violet properties \cite{Horava:2009uw, Blas:2009yd}. Scalar-tensor theories in the unitary gauge (where the scalar is a function of time only) are other interesting examples which have been applied to late time cosmology \cite{Langlois:2015cwa,Langlois:2017mxy,DeFelice:2018mkq}. 

The second remark concerns the relation between the theories \eqref{LQGmod} and effective loop quantum gravity. Once again, let us emphasize that the choice of such theories to model the effects of loop quantum gravity at an effective level is motivated by results 
in symmetry reduced situations where modifications affect the components of the momenta of the metric (see for instance \cite{Bojowald:2004af,Bojowald:2005cb,Modesto:2006mx,Bojowald:2015zha,Corichi:2015xia,Bojowald:2016itl,Gambini:2013ooa,Ashtekar:2018cay}). More rigorous constructions 
of effective theories (based on the full canonical or covariant quantum theory) with no symmetry reduction have been proposed recently and they lead to much more complete and more involved descriptions \cite{Alesci:2018loi,Alesci:2019pbs}.  For the purposes of this paper where we study the possibility to extend the symmetry reduced effective theories to general (non-symmetry reduced) effective theories, the action \eqref{LQGmod} is the most general starting point provided that we require that the modifications are local and affect only the extrinsic curvature.

\subsection{Canonical analysis and deformed Hamiltonian constraint}
The aim of this subsection is to make a canonical analysis of \eqref{LQGmod} to see how the Hamiltonian constraint  is modified compared to 
general relativity. For simplicity, we first assume that the function $F$ in \eqref{funF} depends  on $\text{tr}(K)$ and  $\text{tr}(K^2)$ only. We will discuss later the  general case where $F$ depends also on  $\text{tr}(K^3)$.

We start by introducing the pairs of canonical variables,
\bea
\{ \gamma_{ij}(x),\pi^{kl}(y)\} = \frac{1}{2} (\delta_i^k \delta_j^l +\delta_j^k \delta_i^l  ) \, \delta^3(x-y) \, , \quad 
\{ N(x),\pi_N(y)\} =\delta^3(x-y) \, .
\eea
As the theory is invariant under spatial diffeomorphisms, we do not need to introduce momenta associated to the shift vector $N^i$.
In the deformed theories, the lapse is still not dynamical and we get the primary  constraint $\pi_N = 0$.

An immediate calculation shows that the momenta $\pi^{ij}$ are given by
\bea
\label{momentap}
\pi^{ij} \; \equiv \;\sqrt{\gamma} p^{ij} \; \equiv N \sqrt{\gamma}\frac{\partial F}{\partial \dot \gamma_{ij}} \; = \;  \sqrt{\gamma} \left( \frac{1}{2} F_{(1)} \gamma^{ij}
+ F_{(2)} K^{ij}\right) \, ,
\eea
where $F_{(a)}$ is the partial derivative of $F$ with respect to its first $(a=1)$ or second $(a=2)$ variable. From the expression of the momenta above \eqref{momentap}, it is immediate to see that 
\bea
\label{PitoK}
{\text{tr}(p)} = \frac{3}{2} F_{(1)} + \text{tr}(K) F_{(2)} \, , \quad
{\text{tr}(p^2)} = \frac{3}{4} F_{(1)}^2 + \text{tr}(K^2) F_{(2)}^2 + \text{tr}(K) F_{(1)} F_{(2)} \, ,
\eea
which implies, as expected, that $\text{tr}(p)$ and $\text{tr}(p^2)$ 
are scalars that can be expressed in terms of $\text{tr}(K)$
and $\text{tr}(K^2)$. Locally (if $F$ satisfies regularity conditions which we assume to be true), one can reverse these relations as follows
\bea
\label{KtoPi}
\text{tr}(K) = A \left({\text{tr}(p)} , {\text{tr}(p^2)}\right) \, , \quad
\text{tr}(K^2) = B  \left({\text{tr}(p)} , {\text{tr}(p^2)}\right) \, ,
\eea
where the explicit form of the functions $A$ and $B$ is not needed here. Then, one makes use of these relations to invert the equations between the velocities and the momenta, and to solve $K^{ij}$ in terms of $p^{ij}$,
\bea
K^{ij} = \frac{1}{F_{(2)}} \left( {p^{ij}} - \frac{F_{(1)}}{2} \gamma^{ij} \right) \, .
\eea
In these equations, $F_{(1)}$ and $F_{(2)}$ are viewed as functions of $\text{tr}(p)$ and $\text{tr}(p^2)$ using
\eqref{KtoPi}. Finally, the Hamiltonian can be obtained immediately, and after a short calculation, we show that
\bea
H  & = & \int d^3x \sqrt{\gamma} \left( 
p^{ij} \dot{\gamma}_{ij} - N  \left[ F\left(\text{tr}(K),\text{tr}(K^2)\right) + R \right]  \right) \\
&=& \int d^3x \sqrt{\gamma} \left( 
N {\cal H} + N^i {\cal H}_i 
\right) \, ,
\eea
where ${\cal H}_i \equiv -2 D^j p_{ij}$ is the usual vectorial constraint whereas ${\cal H}$ is a deformed Hamiltonian constraint
given  by
\bea
\label{Hamdef}
{\cal H} \equiv G(\text{tr}(p),\text{tr}(p^2)) - R \, , \qquad \text{where} \quad
G \equiv  \frac{2}{F_{(2)}} {\text{tr}(p^2)} - \frac{F_{(1)}}{F_{(2)}} {\text{tr}(p)} - F  \, .
\eea
Here again, the function $F$ and its derivative are viewed as functions of $\text{tr}(p)$ and $\text{tr}(p^2)$ using
\eqref{KtoPi}. As there are no restrictions on the function $F$ (up to some regularity conditions which allow the inversion \eqref{KtoPi}), 
the function $G$ is also arbitrary, and the action \eqref{LQGmod} leads to a  generic loop-like deformation of the Hamiltonian constraint 
where only the momenta component of the Hamiltonian is affected. Hence, as we already said previously, 
it provides a very general model to test effective loop quantum gravity theories where only the Hamiltonian constraint is modified. 
The theory remains invariant under spatial diffeomorphisms, and the  vectorial constraint is unchanged. 

Notice that, if we had started with a function $F$ which would have depended also on the variable $\text{tr}(K^3)$, we would have obtained an 
expression for the Hamiltonian constraint similar to \eqref{Hamdef} with a function $G$ depending also on $\text{tr}(p^3)$. In the rest of the paper, we still restrict our study to the case  \eqref{Hamdef}. 

\subsection{Deformed Hamiltonian constraint vs. closed algebra of constraints}
\label{sectionFull}
Once the Hamiltonian has been computed, one continues the Hamiltonian analysis to see whether the deformed Hamiltonian constraint is
first or second class. As $\cal H$ is a scalar with respect to spatial diffeomorphisms, it commutes obviously with the vectorial constraint and
studying the stability of the Hamiltonian constraint under time evolution reduces to computing the Poisson algebra
\bea
C[u,v] \equiv \{ {\cal H}[u] , {\cal H}[v]\} \, , \quad \text{with} \qquad {\cal H}[u] \equiv \int d^3x \sqrt{\gamma} \, u \, (G-R) \, ,
\eea
for any functions $u$, $v$ on $\Sigma$. If $C[u,v]$ is weakly vanishing, ${\cal H}$ is first class and the canonical analysis stops here with the conclusion that the theory propagates 2 degrees of freedom as in general relativity. 
On the other hand, when $C[u,v]$ is not weakly vanishing, the theory admits a new 
(secondary) constraint and one has to analyse further the properties of this constraint to know the number of degrees of freedom. 

The calculation of $C[u,v]$ can be done as follows. From the very definition of the Poisson bracket, we have
\bea
\label{Cuv1}
C[u,v] = \int d^3z \left[ \frac{\delta}{\delta \gamma^{ij}(z)} \int d^3x \sqrt{\gamma(x)} u(x) R(x) \right]
\left[ \int d^3y \sqrt{\gamma(y)} v(y) \frac{\delta G(y)}{\delta \pi_{ij}(z)}\right] - (u \leftrightarrow v) \, ,
\eea
where $(u \leftrightarrow v)$ means that we add the same expression where the roles of the functions $u$ and $v$ are inverted, so that
$C[u,v]$ is skew-symmetric. Then, using the definition of the 3-dimensional Ricci scalar, we obtain
\bea
\label{Cuv2}
C[u,v] = \int d^3x \sqrt{\gamma} \left[ D^i D^j u - \gamma^{ij} \Delta u \right] v  \left[ G_{(1)} \gamma_{ij} + 2 G_{(2)} p_{ij} \right] - (u \leftrightarrow v) \, ,
\eea
where  $\Delta = D^i D_i$ is the Laplacian, and $G_{(a)}$ is the derivative of $G$ with respect to its first ($a=1$) or second $(a=2)$
variable. Later, we will also use the notation $G_{(ab)}$ for the second partial derivatives of $G$. 
The initial 3 integrals in \eqref{Cuv1} reduces to only one integral \eqref{Cuv2} after some integrations by part. The rest of the calculation is straightforward and one obtains finally
\bea
\label{Cuv3}
C[u,v] = 2 \int d^3x \sqrt{\gamma} (u D^j v - v D^j u) \left[ G_{(2)} {D^i p_{ij}} + {\cal C}_j \right] \, ,
\eea
where
\bea
\label{DefofC}
{\cal C}_j \equiv p_{ij} \, D^i G_{(2)} - D_j (G_{(1)} + \text{tr}(p) G_{(2)}) \, .
\eea
Thus, $C[u,v]$ reduces to a sum of two terms in the last parantheses \eqref{Cuv3}. The first one is nothing but the vectorial constraint which is obviously weakly vanishing. In general, the second term \eqref{DefofC} is not vanishing and leads to new constraints in the theory. Before discussing these
new constraints, let us ask the question whether one can get a closed algebra of constraints with a deformed Hamiltonian constraint in 
a non-symmetry reduced case. This happens only when ${\cal C}_j$ vanishes (at least weakly). An immediate calculation shows that
${\cal C}_j$ can be written as a sum of four terms,
\bea
{\cal C}_j & = & -[ G_{(11)} + G_{(2)} + \text{tr}(p) G_{(12)}] D_j \text{tr}(p) +  G_{(12)} p_{ij} D^i [ \text{tr}(p)]  \nonumber \\
&&-[G_{(12)} + G_{(22)} \text{tr}(p)] D_j \text{tr}(p^2) +G_{(22)} {p_{ij} D^i [  \text{tr}(p^2)]} \, ,
\eea 
which have independent tensorial structures. 
Hence, ${\cal C}_j$ vanishes only if each term vanishes independently, which leads to the
following four conditions of the function $G$,
\bea
G_{(11)} + G_{(22)} + \text{tr}(p) G_{(12)} =0 \, , \quad
 G_{(12)} + \text{tr}(p) G_{(22)} =0 \, , \quad
  G_{(12)} =  G_{(22)} =0 \, .
\eea
The general solution, which can be computed immediately, reads (up to a global constant factor)
\bea
G = \text{tr}(p^2) - \frac{1}{2} \text{tr}(p)^2 + \alpha \text{tr}(p) + \lambda \, ,
\eea 
where $\alpha$ and $\lambda$ are constant. Without loss of generality, one can always fix $\alpha$ to zero from a canonical transformation
of the form $p_{ij} \mapsto p_{ij} + \beta \gamma_{ij}$ whereas $\gamma_{ij}$ is unchanged. In that case, we recover the Hamiltonian constraint 
of general relativity supplemented with a cosmological constant.

As a consequence, none of the theories in the class \eqref{LQGmod}, which is
different from general relativity with a cosmological constant, admits a closed algebra of constraints generated by the 
usual vectorial constraints and a deformed Hamiltonian constraint. Furthermore, the stability under time evolution of the deformed 
Hamiltonian constraint leads to  a differential equation for the lapse function,
\bea
{\cal S} \equiv N D^i {\cal C}_i + 2 {\cal C}_i D^i N \; \approx \; 0 \, ,
\eea
which has to be understood as a new constraint in the theory. We are using the standard notation $\approx$ for the weak
equality (equality on the constraints surface).
Not only $\cal S$ does not commute with $\pi_N$,
but also their Poisson bracket is non-local in the sense that it involves derivatives of delta distributions. Such a situation is pathological
and makes the theory ill-defined with an undefined number of degrees of freedom. 

\medskip

The conclusion of this analysis is that one cannot extend the condition of having a closed algebra of constraints associated to an 
effective loop quantum gravity theory for an arbitrary (non-symmetry reduced) background which satisfies the following properties: first, it is  invariant under spatial diffeomorphisms; second, the deformation does not involve the 3-dimensional Ricci tensor but only the extrinsic curvature tensor; and finally the effective theory is a local theory of the metric. 
One should relax one of these hypothesis to construct an effective theory of loop quantum gravity. As the fundamental variables in loop quantum gravity are holonomies of a connection, one may expect to get instead 
a non-local  or a non-metric deformation.

Nonetheless, requiring a closed algebra of constraints for an effective theory of loop quantum gravity (satisfying the previous properties) 
 for spherically symmetric space-times becomes possible for some reasons we are going to explain later. 
 Their constructions lead  to very interesting scenarii, as we are going to see in the next two sections.

\section{Deformations of Spherically Symmetric Gravitation}
\label{sectionIII}

From now on, we restrict our study to (non-static) spherically symmetric space-times. Even if the deformed gravity theory in this symmetry reduced sector can not be embedded into a fully covariant theory, it is still of interest to consider such specific models for at least two reasons. First, it provides a rare enough example of a gravitational system with deformed covariance, and leads to interesting peculiar effects, such as transition between lorentzian and euclidean regimes in the deep interior for which we have explicit solutions. Second, this deformed covariance is not tied to spherical symmetry, but is also realized in more general system, such as cylindrical symmetry reduced gravity (or Gowdy systems). Hence, while deformed covariance can only be realized in symmetry reduced systems, it provides an interesting testbed to understand the spacetime description beyond standard GR.

Focusing therefore on the spherically symmetric sector, one can choose a coordinate system such that 
the  spatial line element $d\ell^2$ takes the simple form
\bea
\label{sphericalmetric}
d\ell^2 \equiv \gamma_{ij} dx^i dx^j = \gamma_{rr} dr^2 + \gamma_{\theta \theta} (d\theta^2 + \sin^2\theta d\varphi^2) \, ,
\eea
where $\gamma_{rr}$ and $\gamma_{\theta\theta}$ are functions of $(t,r)$ only. We also assume that the lapse function $N$ and the (radial
component of the) shift vector $N^r$ are functions of $(t,r)$ only.

We start this section with a quick review of the Hamiltonian analysis of general relativity reduced to spherical symmetry. This will give us the opportunity to introduce some notations. Afterwards, we will study the possibility to deform  the theory \`a la loop quantum gravity. We will adopt a Hamiltonian point of view and ask the question of the possibility to deform the constraints keeping a closed Poisson algebra. 
First, we will prove the impossibility to
deform the vectorial constraint without introducing anomalies (in the algebra of constraints). However, as it is well known, we will show on the other hand that
it is possible to deform the Hamiltonian constraint keeping a closed algebra, and we will classify these deformations. This results contrasts 
with the results we have obtained in the previous section where we have shown the impossibility to deform (with some hypothesis) the Hamiltonian constraint for a generic background with no particular symmetries. We will discuss why reducing to spherical symmetry makes
the deformation possible.

\subsection{Reduction to spherical symmetry}
\label{spherical}
We start with a review of the Hamiltonian analysis of general relativity for non-static spherically symmetric geometries. 
The ADM parametrization of the metric is given by \eqref{metric_ADM} where $N^r$ is the only non-trivial component of the shift vector 
and the spatial metric is \eqref{sphericalmetric}. The only non-vanishing components of the extrinsic curvature \eqref{Eij} are
\bea
K_r^r & = &\frac{1}{2N} \left( \frac{\dot \gamma_{rr}}{\gamma_{rr}} - N_r \frac{\partial_r \gamma_{rr}}{\gamma_{rr}^2} - 2 \frac{\partial_r N_r}{\gamma_{rr}}\right) \, , \\
K_\theta^\theta & = & K_\varphi^\varphi = \frac{1}{2N} \left( \frac{\dot \gamma_{\theta\theta}}{\gamma_{\theta\theta}} + N_r \frac{\partial_r \gamma_{\theta\theta}}{\gamma_{rr}\gamma_{\theta\theta}}\right) \, .
\eea
Therefore, the Einstein-Hilbert action \eqref{EHaction} reduces to (up to an irrelevant global constant that we neglect but
can be computed from the integration  over the angles $\theta$ and $\varphi$)
\bea
\label{spheraction}
S_{EH} =  \int   dr \, dt \,  L_{EH}  = 
\int dr \, dt \, N \sqrt{\gamma_{rr}} \gamma_{\theta\theta} \left( -2 (K_\theta^\theta)^2 - 4 K_\theta^\theta K_r^r + R\right) \, .
\eea
We recall that $R$ is the 3-dimensional Ricci scalar for  spherically symmetric background whose expression will be given later on.

To go further, it is convenient to introduce the ``electric fields'' $E^r$ and $E^\varphi$ associated to the metric \eqref{sphericalmetric} which are defined by the relations,
\bea
\label{metric coeff}
\gamma_{rr} \equiv \frac{(E^\varphi)^2}{E^r} \, , \qquad \gamma_{\theta \theta} \equiv E^r \, ,
\eea
with the condition that $E^\varphi > 0$. Hence,  \eqref{spheraction} is now considered as an action for the dynamical variables $E^r$ and
$E^\varphi$. To perform its canonical analysis, one introduces the two pairs of conjugate momenta,
\bea
\label{PSvar}
\{ E^r (x) , \pi_r(y)\} = \delta(x-y) \, , \qquad
\{ E^\varphi (x) , \pi_\varphi(y)\} = \delta(x-y) \, ,
\eea
whereas $N$ and $N^r$ are considered as Lagrange multipliers. Notice that the notations $x$ and $y$ refer to the radial coordinate. The momenta are easily computed and read 
\bea
\pi_\varphi = - 4 \sqrt{E^r} K_\varphi^\varphi \, , \qquad
\pi_r = -  2 \frac{E^\varphi}{2\sqrt{E^r}} K_r^r \, .
\eea
Then we compute the Hamiltonian of the theory,
\bea
\label{Hamiltonian}
H \equiv \int dr  \,  \left( \pi_r \dot{E}^r + \pi_\varphi \dot{E}^\varphi - L_{EH} \right)=  \int dr \, \left(  N {\cal H} + N^r {\cal H}_ r \right)
\eea
where the Hamiltonian and vectorial constraints are respectively given by,
\bea
{\cal H} \equiv -\frac{E^\varphi}{2 \sqrt{E^r}} \pi_\varphi^2 - 2{\sqrt{E^r}}\pi_\varphi \pi_r - 4{E^\varphi }\sqrt{E^r} R \, , \quad
{\cal H}_r \equiv E^\varphi \pi_\varphi' - \pi_r (E^r)'  \, ,\label{classical constraints}
\eea
up to a rescaling of the lapse function, with the following expression of the Ricci scalar,
\bea
4R =  \frac{1}{2E^r} - \frac{((E^\varphi)')^2}{8 (E^\varphi)^2 E^r} + \frac{(E^r)' (E^\varphi)'}{2(E^\varphi)^3} - \frac{(E^r)''}{2(E^\varphi)^2}  \, .
\eea
We are using the notation $f'$ for the derivative of any function $f$ with respect to $r$. 

It is well-known that $\cal H$ and ${\cal H}_r$ define first class constraints, they generate the invariance under diffeomorphisms
for non-static spherically symmetric backgrounds, and they satisfy the closed Poisson algebra,
\bea
\{ {\cal H}_r[u] , {\cal H}_r[v]\}  & = & {\cal H}_r[u'v-uv'] \, , \label{HrHr}\\
\{ {\cal H}_r[u] , {\cal H}[v]\}  & = &  - {\cal H}[u v' ] \, ,  \label{HrH}\\
\{ {\cal H}[u] , {\cal H}[v]\}  & =  & {\cal H}_r[\gamma^{rr}(uv'-vu')] \, . \label{HH}
\eea
where $ {\cal H}_r[u]$ and  ${\cal H}[u]$ are the smeared constraints, $u$ being a regular function of $r$. 
As a consequence, the Dirac analysis of the constraints stops here, and the theory propagates no degrees of freedom.

\medskip

One expects loop quantum gravity to modify, at the effective level, the Einstein equations. 
In the Hamiltonian framework, this means that one expects modifications of the constraints of the theory.  
Even though we have shown in the previous section that it is not possible to obtain a non-trivial effective deformation 
of the full theory with local modifications of the Einstein-Hilbert action (of the form \eqref{LQGmod}) which keep the constraints algebra
closed, we can ask the same question when we consider spherically symmetric space-times. It is well-known that this problem has a solution in that case.  Now, we are going to derive  the conditions for  deformed  vectorial and Hamiltonian constraints to still have a closed Poisson algebra. 

\subsection{Deformation of the vectorial constraint}
\label{DefofDiff}
From the construction of  kinematical states in loop quantum gravity, we know that one has to keep the invariance under spatial diffeomorphisms, or at least to keep a closed algebra for an eventual deformation of the vectorial constraint (to avoid having anomalies). 
Indeed, spatial diffeomorphisms are still symmetries of the kinematical Hilbert space of loop quantum theory, and there is no reason to violate
such a symmetry at the effective level. However, one can wonder if the (spatial) diffeomorphism algebra could be eventually deformed (compared to the classical case)  but still closed. We are going to explore this problem and  show that, under some general hypothesis, 
this is not possible.

\subsubsection{First necessary condition}
For that purpose, we start with the simple remark that $\pi_\varphi$ and $E^r$ transform as scalars under the action of the vectorial constraint
${\cal H}_r$ whereas $E^\varphi$ and $\pi_r$ are densities. Hence, for convenience (in this subsection only), we introduce the notations,
\bea
q_1 \equiv \pi_\varphi \, , \quad p_1 \equiv - E^\varphi \, , \quad q_2 \equiv E^r \, , \quad p_2 \equiv \pi_r \, ,
\eea  
so that 
\bea
\{q_i(x),p_j(y)\} = \delta_{ij} \delta(x-y) \, , \qquad {\cal H}_r= - (p_1 q_1' + p_2 q_2') \,.  
\eea
From the point of view of the vectorial constraint, the two degrees of freedom $(q_1,p_1)$ and $(q_2,p_2)$ are decoupled {(there are no cross terms involved)}, and we look 
for modifications which conserve this decoupling. 

Hence, we ask the question whether there exists a function ${\cal D}(q,q',p,p')$ such that the deformed vectorial constraint,
\bea
\label{D1D2}
{\cal H}_{r,\rm{def}} \equiv {\cal D}(q_1,q_1',p_1,p_1') +{\cal  D}(q_2,q_2',p_2,p_2' ) \, \approx \, 0 \, ,
\eea
satisfies a closed Poisson algebra. We can treat the two components separately and compute the Poisson brackets
between the smeared function
\bea
{\cal D}[u] \equiv \int dr \, u(r) \,  {\cal D}(q,q',p,p') \, , 
\eea
where we omit, for simplicity, the labels ($1$ or $2$) for the position and momentum variables. Variations of ${\cal D}[u]$ induced by variations
$\delta q$ and $\delta p$ of $p$ and $q$ respectively are easily computed and read
\bea
\delta {\cal D}[u] \; = \; \int dr \, \left[ 
\delta q \, (u {\cal D}_q -(u {\cal D}_{q'})') + \delta p  \, (u {\cal D}_p -(u {\cal D}_{p'})')
\right] \, ,
\eea
where ${\cal D}_z$ denotes the derivative of ${\cal D}$ with respect to $z \in \{q,q',p,p' \}$. Hence,  after an immediate calculation,  
we show that
\bea
\label{DD1}
\{ {\cal D}[u]  , {\cal D}[v] \} & = & \int dr \, (uv'-u'v)  \left[ {\cal D}_q {\cal D}_{p'} - {\cal D}_{q'} {\cal D}_p + {\cal D}_{q'} ({\cal D}_{p'q} q' + {\cal D}_{p'p} p' + {\cal D}_{p'q'}q'' + {\cal D}_{p'p'} p'') \right. \nonumber \\
&& \hspace{3cm} - \left. {\cal D}_{p'} ({\cal D}_{q'q} q' + {\cal D}_{q'p} p' + {\cal D}_{q'q'} q'' + {\cal D}_{q'p'} p'')\right] \, .
\eea
For the algebra to be closed, the terms proportional to $p''$ and $q''$ in \eqref{DD1} must vanish, which implies the two necessary conditions
\bea
{\cal D}_{q'} {\cal D}_{p'q'} - {\cal D}_{p'} {\cal D}_{q'q'} = 0 = {\cal D}_{q'} {\cal D}_{p'p'} - {\cal D}_{p'} {\cal D}_{q'p'} \, .
\eea
These two equations are obviously similar and then are solved in the same way. we concentrate on the first one which is solved as follows:
\bea
\frac{{\cal D}_{q'q'}}{{\cal D}_{q'}} - \frac{{\cal D}_{q'p'}}{{\cal D}_{p'}} = 0 \Longleftrightarrow \frac{\partial }{\partial q'} \left( \ln {\cal D}_{q'} - \ln {\cal D}_{p'}\right) = 0 
\Longleftrightarrow   {\cal D}_{p'}=A(q,p,p')  {\cal D}_{q'}  \, ,
\eea 
where the function $A$ is arbitrary. Similarly, the second condition leads to
\bea
{\cal D}_{p'} = B(q,p,q') {\cal D}_{q'} \, ,
\eea
where  the function $B$ is also arbitrary. As a consequence, $A(q,p,p') =B(q,p,q')=C(q,p) $ and  the general solution of the two previous
conditions is
\bea
\label{DefDif}
{\cal D}(q,q',p,p') = D\left(q' + C(q,p) p'\right) \, ,
\eea
where $D$ is an arbitrary function of one variable. Hence, any deformation of the diffeomorphisms algebra is necessarily of the form \eqref{DefDif}. 
Before going further and solving the remaining conditions, we make a canonical transformation that will drastically simplifies the analysis. 

\subsubsection{Canonical transformation: simplification of the problem}
Indeed, we introduce a new pair of canonically conjugate variables $(Q,P)$ related to $(q,p)$ via the generating function $\Phi(p,Q)$ by the relations
\bea
q = \frac{\partial \Phi}{\partial p} \equiv \alpha(p,Q) \, , \qquad 
P= \frac{\partial \Phi}{\partial Q}\equiv \beta(p,Q) \, .
\eea 
Hence, the combination $q' + C(q,p) p'$ that appears in \eqref{DefDif}  transforms under this canonical transformation according to
\bea
q' + C(q,p) p' = \alpha_Q Q' + \left[ \alpha_p + C(\alpha(p,Q),p)\right] p' \, ,
\eea
and then we can always choose a generating function $\Phi$  such that its partial derivative
$\alpha$ satisfies the condition
\bea
\label{pdealpha}
\alpha_p (p,Q) + C\left( \alpha(p,Q),p\right) = 0 \, ,
\eea
so that $q' + C(q,p) p'= \alpha_Q Q'$ does not depend on $P'$, and the deformed constraint takes the simple form
\bea
{\cal D}(q,q',p,p') = D(\alpha_Q(Q,P) Q') \, ,
\eea
where we used the shorthand $\alpha_Q(Q,P)$ for $\alpha_Q (p(Q,P),Q)$.  When the function $C$ is regular enough, the first
order partial differential equation \eqref{pdealpha} admits always a solution at least locally.

As a consequence, up to canonical transformations,  the deformed constraint can be reduced, without loss of generality, to the simple form
\bea
\label{SimplifiedDefDif}
{\cal D}(q,q',p,p') \; = \; D(X) \, , \qquad X \equiv J(q,p)q' \, ,
\eea
where $J$ is an arbitrary function. 

\subsubsection{No-go: no closed algebra for deformed diffeomorphisms constraints}
At this stage, we return to the expression of the Poisson bracket \eqref{DD1} between the deformed constraints and we immediately see 
that it simplifies considerably when ${\cal D}$ is of the form \eqref{SimplifiedDefDif},
\bea
\{ {\cal D}[u]  , {\cal D}[v] \} & = & \int dx \, (u'v-uv')  {\cal D}_{q'} {\cal D}_p \, .
\eea
Hence, the full constraint \eqref{D1D2} has a closed Poisson algebra if and only if 
\bea
{\cal D}_{q'} {\cal D}_p = \lambda {\cal D} \, ,
\eea
where $\lambda$ is necessarily a constant (independent of $q_1$, $p_1$, $q_2$, $p_2$ and their derivatives). Substituting
\eqref{SimplifiedDefDif} into this equation leads to 
\bea
\frac{X D_X^2}{D} = \frac{\lambda}{J_p} \, ,
\eea
which can be solved easily for the function $D$ that is given by
\bea
D = \lambda \frac{X}{J_p} \, .
\eea
As $D$ is a function of $X$ only by definition, then $J_p$ is necessarily a constant. Finally, we arrive at the conclusion that the only
constraints \eqref{D1D2} which admits a closed Poisson algebra are such that
\bea
{\cal D}(q,q',p,p') = \lambda (p + \mu(q)) q' \, ,
\eea
where we recall that $\lambda$ is a constant, and the arbitrary function $\mu(q)$ can be set to zero without loss of generality 
by a simple canonical 
transformation. Hence, there is no deformation of the diffeomorphisms Poisson algebra which preserves the decoupling \eqref{D1D2}. 

\subsection{Deformation of the scalar constraint}
Now, we review the possibility to deform the Hamiltonian constraint keeping a closed (but deformed) constraints Poisson algebra.
Hence, we look for deformed Hamiltonian constraints ${\cal H}_{\rm{def}}$, which are functions on the phase space, such that
\eqref{HrH} remains unchanged and \eqref{HH} is deformed but closed. 

\subsubsection{General point-holonomy deformation of the scalar constraint}
First of all, the fact that the Poisson bracket \eqref{HrH} is unchanged  implies that the deformed Hamiltonian 
constraint ${\cal H}_{\rm{def}}$ is a scalar density of weight of +1.  Furthermore, as we are considering point-holonomy corrections only, 
the Ricci part of the classical Hamiltonian constraint \eqref{classical constraints} is unchanged and then ${\cal H}_{\rm{def}}$ can be written 
in the form
\bea
\label{deformed H}
{\cal H}_{\rm{def}} \; = \; \sqrt{\gamma} \, \left[ S(\pi_\varphi, \frac{\pi_r}{E^\varphi},E^r) - 4 R \right]\, ,
\eea
where $S$ is an arbitrary function of the three independent scalars,
\bea
\label{Xvar}
X_1 \equiv \pi_\varphi \, , \quad X_2 \equiv \sqrt{E^r}\frac{\pi_r}{\sqrt{\gamma}}= \frac{\pi_r}{E^\varphi}\, , \quad X_3 \equiv E^r \, ,
\eea
which generate the algebra of scalar functions in the phase space.
Notice that the  factor $4$ that comes with the Ricci scalar $R$ has been introduced to make the comparisons with the classical constraint \eqref{classical constraints} easier.

Now, we compute the Poisson bracket between two deformed Hamiltonian constraints \eqref{deformed H}. 
For simplicity, we introduce the notation
\bea
\label{NewfuncG}
G(\pi_\varphi,\pi_r,E^\varphi,E^r) \equiv \sqrt{\gamma} \, S(\pi_\varphi, \frac{\pi_r}{E^\varphi},E^r) \, ,
\eea
and we show that, for any pair of functions $(u,v)$, the Poisson bracket between the smeared deformed constraints ${\cal H}_{\rm{def}}[u]$
and ${\cal H}_{\rm{def}}[v]$ is given by
\bea
\{{\cal H}_{\rm{def}}[u],{\cal H}_{\rm{def}}[v] \} = \int dr \, ds \,  \left[ u(r) v(s) - u(s) v(r)\right] \left\{ G(r) , - \sqrt{h(s)} R(s) \right\} \,,
\eea
which, after a long but straightforward calculation, reduces to
\bea
\{{\cal H}_{\rm{def}}[u],{\cal H}_{\rm{def}}[v] \} = \int dr \, ( u v' - u' v) \frac{\sqrt{E^r}}{2 E^\varphi} \left[ \frac{(E^r)'}{E^\varphi} 
\frac{\partial G}{\partial \pi_\varphi} - \left( \frac{\partial G}{\partial \pi_r}\right)' \right] \, .
\eea
If one substitutes the expression of $G$ in terms of the function $S$ \eqref{NewfuncG}, this Poisson bracket becomes
\bea
\label{HdefHdef}
\{{\cal H}_{\rm{def}}[u],{\cal H}_{\rm{def}}[v] \} & = & \int dr \, ( u v' - u' v) \frac{\sqrt{E^r}}{2 E^\varphi} \left[  
(E^r)' \sqrt{E^r} S_{(1)} - \frac{(E^r)'}{2 \sqrt{E^r}} S_{(2)} \right. \nonumber \\
&& \;\;\;\;\;\;\;\;\;\;\;\;\;\;\;\;\;\;\;\;\;\;\; \left.- \sqrt{E^r} \left( S_{(12)} \pi_\varphi' + S_{(22)} \left( \frac{\pi_r}{E^\varphi}\right)'
+ S_{(23)} (E^r)' \right) 
\right] \, ,
\eea
where $S_{(a)} \equiv \partial S/\partial X_a$ and $S_{(ab)} \equiv \partial^2 S/(\partial X_a \partial X_b)$ denote the partial derivatives of the function $S$ with respect to the variables $X_a$ \eqref{Xvar}.

\subsubsection{Closeness of the deformed algebra}
The constraints algebra is closed if and only if
\bea
\{{\cal H}_{\rm{def}}[u],{\cal H}_{\rm{def}}[v] \} & = & {\cal H}_r[\gamma^{rr}_{\rm def} ( u v' - u' v)]  \, ,
\eea
where $\gamma^{rr}_{\rm def}$ could be an arbitrary function on the phase space which represents the deformation of the (inverse) metric component 
$\gamma^{rr}$.
 As a consequence, from \eqref{HdefHdef}, we immediately see that 
a necessary condition for the algebra to be closed is that $S_{(22)}=0$ which implies that $S$ is an affine function of $X_2$
and then, as $\sqrt{\gamma} = E^\varphi \sqrt{E^r}$, 
\bea
\label{SAB}
S(\pi_\varphi,\frac{\pi_r}{E^\varphi},E^r) = A(\pi_\varphi,E^r) + B(\pi_\varphi,E^r) \frac{\pi_r}{E^\varphi} \, ,
\eea
where $A$ and $B$ are arbitrary functions.
In that case, it is easy to see that \eqref{HdefHdef} reduces to
\bea
\{{\cal H}_{\rm{def}}[u],{\cal H}_{\rm{def}}[v] \}  & = & {\cal H}_r[ \gamma^{rr}_{\rm def} ( u v' - u' v) ]  \nonumber \\
&+& \int dr \, ( u v' - u' v)  \frac{E^r (E^r)' }{2 E^{\varphi}} 
\left[ \frac{\partial A}{\partial \pi_\varphi} - \frac{\partial B}{\partial E^r} - \frac{B}{2 E^r} \right] \, ,
\eea
with the deformed (inverse) metric coefficient given by 
\bea
\label{def inv metric}
\gamma^{rr}_{\rm def}  \equiv - \frac{{E^r}}{2(E^\varphi)^2} \frac{\partial B}{\partial \pi_\varphi} \, .
\eea
As a consequence, the algebra is closed if and only if the following condition is satisfied
\bea
\label{close cond}
 \frac{\partial A}{\partial \pi_\varphi} - \frac{\partial B}{\partial E^r} - \frac{B}{2 E^r} \; = \; 0 \; ,
\eea
otherwise there is an anomaly and the Hamiltonian constraint is no more first class, which means that it does not generate any symmetries.
This conclusion is consistent with what has already been found in the literature \cite{Tibrewala:2012xb,Bojowald:2015zha,Bojowald:2016hgh,Bojowald:2016itl}.  

Before going further and discuss this condition, notice that in general relativity the Hamiltonian constraint is of the form \eqref{deformed H}
with \eqref{SAB} where the functions $A$ and $B$ are explicitly given by
\bea
A \; = \; -\frac{\pi_\varphi^2}{2 E^r} \, , \qquad
B \; = \; - 2 \pi_\varphi \,,
\eea
which obviously satisfy the closeness condition \eqref{close cond}.  Furthermore, the deformed inverse metric \eqref{def inv metric} 
reduces, in that case, to the classical inverse metric $\gamma^{rr}$ \eqref{metric coeff}.

\subsubsection{Discussion}
We found deformations of spherically symmetric reduced general relativity that lead to a closed constraints algebra where the vectorial constraint
is unchanged but the Hamiltonian constraint is deformed by (point-) holonomy like corrections. These theories admit two different symmetries 
which are the classical invariance under diffeomorphisms and an invariance under a deformed time reparametrisation. Hence, they have no local degrees of freedom, exactly as general relativity when it is reduced to spherical symmetry. The deformed symmetry has been discussed
recently in \cite{Bojowald:2018xxu}.

The possibility to have a closed (deformed) constraints algebra relies on the fact that we are considering spherically
symmetric backgrounds only. The reason is simply that there is only one spatial diffeomorphisms constraint in the symmetry reduced theory whereas there are obviously three of them in the full theory. As we showed in section \eqref{sectionFull}, requiring the invariance under
3 dimensional spatial diffeormorphisms leads to no-go results and to the impossibility of having a non-trivial Hamiltonian (local and point-holonomy like) deformation of general relativity. 

To illustrate better the specificity of spherically symmetric models compared to fully (3 dimensional) covariant theories,  let us study the 
spherically symmetric reduced version of \eqref{LQGmod} when $F$ is of the form \eqref{funF} with no $\text{tr}(K^3)$ dependency. 
 For that purpose, we use the
same notations as in section \eqref{spherical} and we compute the momenta canonically conjugated to $E^r$ and $E^\varphi$ which are
given by the relations
\bea
\frac{E^r \pi_r}{\sqrt{\gamma}} = \frac{1}{2} F_{(1)} + (2 K_\varphi^\varphi - K_r^r) F_{(2)} \, , \qquad
\frac{E^\varphi \pi_\varphi}{\sqrt{\gamma}} = F_{(1)} + 2 K_r^r F_{(2)} \, ,
\eea
where $F_{(1)}$ and $F_{(2)}$ are the derivatives of $F$ with respect to $\text{tr}(K)$ and $\text{tr}(K^2)$ respectively. From these expressions, we immediately show the two following  relations
\bea
&&\frac{E^r \pi_r }{\sqrt{\gamma}} + \frac{ E^\varphi \pi_\varphi }{\sqrt{\gamma}} = \frac{3}{2} F_{(1)} + \text{tr}(K) F_{(2)} \, , \\
&&\left(\frac{E^\varphi \pi_\varphi}{\sqrt{\gamma}}\right)^2 + 2 \left( \frac{E^r \pi_r }{\sqrt{\gamma}} + \frac{1}{2}\frac{ E^\varphi \pi_\varphi}{\sqrt{\gamma}} \right)^2
= 2 F_{(1)}^2 + 4 \text{tr}(K) F_{(1)} F_{(2)} +
4 \text{tr}(K^2) F_{(2)}^2 \, ,
\eea
which imply that  $\text{tr}(K)$ and $\text{tr}(K^2)$ can be (at least implicitly) expressed in terms of the combinations ${E^r \pi_r}/{\sqrt{\gamma}}$ and ${E^\varphi \pi_\varphi}/{\sqrt{\gamma}}$. As a consequence, one shows after a long but straightforward calculation
that the Hamiltonian is of the usual form \eqref{Hamiltonian} with a classical vectorial constraint and a deformed Hamiltonian constraint 
\eqref{deformed H} with
\bea
S(\pi_\varphi,\frac{\pi_r}{\sqrt{\gamma}},E^r) = \tilde{S} \left( \frac{E^r \pi_r}{\sqrt{\gamma}} ,\frac{E^\varphi \pi_\varphi}{\sqrt{\gamma}}\right)  \equiv
F_{(1)} \text{tr}(K) + 2 F_{(2)} \text{tr}(K^2) \, ,
\eea
where, in the last equation, $\text{tr}(K)$ and $\text{tr}(K^2)$ are expressed in terms of  
\bea
\frac{E^r \pi_r}{\sqrt{\gamma}} = E^r  \frac{\pi_r}{\sqrt{\gamma}} \, , \qquad
 \frac{E^\varphi \pi_\varphi}{\sqrt{\gamma}} =  \frac{\pi_\varphi}{\sqrt{E^r}} \, .
\eea
As a consequence, these theories have a deformed Hamiltonian constraint of the type \eqref{deformed H}. Hence, they fall in the class 
of theories with a closed deformed constraints algebra if the conditions \eqref{SAB} and \eqref{close cond} are fulfilled even though its fully covariant constraints algebra has an anomaly. Thus, this analysis illustrates clearly that a fully covariant theory with a non-closed constraints 
algebra can lead, in the spherical symmetric sector, to a reduced theory with a closed constraints algebra. This remark raises the possibility
to extend the closeness property beyond spherical symmetry in a fully covariant effective (local and metric) action for loop quantum gravity. 

\subsubsection{The case of a non-closed constraints algebra}
When the conditions \eqref{SAB} and \eqref{close cond} for having a closed constraints algebra are not fulfilled, then the conservation under
time evolution of the deformation Hamiltonian constraint leads to a new constraint given by,
\bea
\label{secondary}
{\cal C} \equiv (E^r)' \sqrt{E^r} S_{(1)} - \frac{(E^r)'}{2 \sqrt{E^r}} S_{(2)} - \sqrt{E^r} \left( S_{(12)} \pi_\varphi' + S_{(22)} \left( \frac{\pi_r}{E^\varphi}\right)'
+ S_{(23)} (E^r)' \right) \approx 0 \, ,
\eea
which reduces to the constraint
\bea
\label{condAetB}
\frac{\partial A}{\partial \pi_\varphi} - \frac{\partial B}{\partial E^r} - \frac{B}{2 E^r} \; \approx \; 0 \; ,
\eea
when only the condition \eqref{SAB} is satisfied and $(E^r)' \neq 0$. 
Generically, the new constraint $\cal C$ has a non-vanishing Poisson bracket with 
the deformed Hamiltonian constraint,
\bea
\{ {\cal H}_{\rm def}[u] , {\cal C}\} \neq 0 \, ,
\eea
for any non-vanishing function $u$.
 As a consequence, in that situation, the deformed Hamiltonian constraint and the secondary constraint 
\eqref{secondary} form a pair of second class constraints. There is no more invariance under deformed time reparametrization and the 
lapse function is a Lagrange multiplier which is fixed by the requirement that the secondary constraint has to be stable under time
evolution. Hence, the equation satisfied by the lapse is simply given by
\bea
\dot{\cal C} = \{ {\cal C}, {\cal H}_{\rm def}[N] \} = 0 \, ,
\eea
which is, in general, a partial differential equation. Clearly $N=0$ is a solution, but it is physically unacceptable. A necessary condition
for the theory to be physically relevant is the existence of a non-vanishing solution for the lapse function.

As in the previous case, these theories do not have local degrees of freedom because
one is replacing the first class constraint ${\cal H}_{\rm def}$ by a pair of second class constraints $({\cal H}_{\rm def},{\cal C})$. Let us emphasize
again that, in this case, the lapse function is not free and can be, in principle, expressed in terms of the phase space variables. 
Contrary to the fully covariant case where the non-closeness of the constraints algebra leads to inconsistencies, 
spherically symmetric models with a non-closed constraints algebra is well-defined, even though it does not exhibit anymore any 
deformed reparametrization invariance. Furthermore, it is not clear whether these theories have, in general, physically relevant solutions.

\section{Effective Black Holes Interior Solutions}

\label{sectionIV}

The purpose of this section is to study (interior) black hole solutions of the theory defined by a deformed Hamiltonian constraint of the form
\eqref{deformed H} while the vectorial constraint is unchanged.  First, we will compute the equations of motion. Then, we will show how to resolve them in full generality. Finally, we will give some concrete examples, and we will compare with solutions already
found in the literature. 

\subsection{Equations of motion}

Here, we look for static spherically symmetric 
black hole solutions where the metric components \eqref{sphericalmetric}, the lapse and the shift vector depend on the ``radial''
 coordinate only. Furthermore, as the deformations of general relativity are  induced by quantum gravity, 
 we are interested in the geometry inside the black hole where quantum  gravity effects are supposed to become important (at least close
enough  to the classical singularity). In the black hole interior (behind the horizon), the radial coordinate plays the role of time and we will 
denote it $t$ instead of $r$. In other words, the role of these variables changes when one crosses the horizon, this corresponds to considering time-dependent {degrees of freedom} only. 

Thus, the equations of motion and the constraints reduce to ordinary differential equations involving time derivatives only.
Hence, the vectorial constraint is trivially satisfied and one can formally forget about the closedness property of the constraints algebra.
Furthermore, the secondary constraint $\cal C$ itself, introduced in \eqref{secondary}, is also trivially satisfied when radial derivatives vanish, and then the lapse function is free.

\medskip

Now, let us  compute the equations of motion. For that, we recall that the time evolution of any function $\cal O$ on the phase space is defined from the Poisson bracket by
\bea
\dot{\cal O} = \{ {\cal O}, {\cal H}_{\rm def}[N] + {\cal H}_r[N^r] \} \, .
\eea
Hence, the equations of motion for the phase space variables	\eqref{PSvar} are easily obtained, and they are partial
differential equations that we cannot solve in full generality. As we said while introducing this section, we restrict our study to
time dependent solutions only. Therefore, one can drop all radial dependency, and the Hamiltonian constraint together with the effective Einstein equations dramatically simplify. Indeed, the Hamiltonian constraint becomes
\bea
\label{Hamcon}
{\cal H}_{\rm def} \; = \; \sqrt{E^r} E^\varphi S \left( \pi_\varphi,\frac{\pi_r}{E^\varphi}, E^r\right) - \frac{E^\varphi}{2 \sqrt{E^r}} \, ,
\eea
and the Einstein equations are
\bea
\dot{E}^r & = & N \sqrt{E^r} S_{(2)} \, ,\label{Ereq} \\
\dot{E}^\varphi & = & N \sqrt{E^r} E^\varphi S_{(1)} \, , \label{Ephieq}\\
\dot{\pi}_r & \approx & - N\sqrt{E^r} E^\varphi \left( S_{(3)} + \frac{1}{2 (E^r)^2}\right) \, ,\label{pireq}\\
\dot{\pi}_\varphi & \approx & \frac{N \sqrt{E^r} \pi_r}{E^\varphi} S_{(2)} \, , \label{piphieq}
\eea
where the weak equality $\approx$ means an equality up to a term proportional to the constraint ${\cal H}_{\rm def}$. 
As the dynamical system reduces to a classical mechanical system (and not a classical field theory because the radial dependency has 
disappeared), the time evolution of the deformed Hamiltonian constraint is trivially satisfied, which can be verified explicitly. 
As a consequence, one can forget about the secondary constraint \eqref{secondary}  and still have a consistent dynamical system.
Indeed, one immediately sees that \eqref{secondary} is trivially satisfied when radial derivatives are vanishing. 

\subsection{General solution of the equations of motion}
Let us show how to solve these equations in full generality. What makes the system ``solvable'' is that it admits a ``triangular" structure in the sense that one can first decouple the variable $E^r$ from the other variables, then solve the equation for $E^r$, and then successively  decouple the equations for $\pi_\varphi$, $E^\varphi$ and $\pi_r$ that can be solved in principle (at least numerically). This method works 
for any function $S$ provided that $S_{(2)}$ does not vanish and generalizes the results of \cite{BenAchour:2018khr}.

\subsubsection{Resolution of the system}
As we announced, we start by solving $E^r$. For that, we fix the lapse function by the condition
\bea
\label{lapsechoice}
N S_{(2)} = 2 \, ,
\eea
which is equivalent to change the time variable $t$ into $\tau$ such that $2 d\tau = N S_{(2)} dt$. In that case, the equation for $E^r$
\eqref{Ereq} simplifies and can solved explicitly according to
\bea
\label{Ersolution}
E^r(t) = (t+a)^2 \, ,
\eea
where $a$ is an integration constant that we fix to $a=0$ in order to recover the Schwarzschild solution at the classical limit.

Then, we concentrate on the equation \eqref{piphieq} for $\pi_\varphi$ which becomes,
\bea
\label{eqforpiphi}
\dot{\pi}_\varphi = 2t \frac{\pi_r}{E^\varphi} \, ,
\eea
where, to simplify notations, we replaced the weak equality $\approx$ by a standard equality. To go further and solve this equation,
we use the Hamiltonian constraint \eqref{Hamcon} which enables us to write the relation
\bea
2 t^2 \, S(\pi_\varphi, \frac{\pi_r}{E^\varphi},t^2) = 1\, .
\eea
As $S_{(2)}$ is supposed not to vanish, we can formally (and locally) invert this equation and solve ${\pi_r}/{E^\varphi}$ in terms of $\pi_\varphi$ and $t$ (by virtue of the implicit function theorem) according to,
\bea
\label{implicitP}
\frac{\pi_r}{E^\varphi} = P(\pi_\varphi,t^2) \, ,
\eea
where $P$ is the (implicit) function defined by the relation
\bea
2 t^2 \, S(\pi_\varphi,P(\pi_\varphi,t^2) ,t^2) = 1 \, .
\eea
Interestingly, when the necessary condition \eqref{SAB} to have a closed constraints algebra is satisfied, the function
$P$ can be explicitly computed and it is given by the expression
\bea
P(\pi_\varphi,t^2) \; = \; \frac{1 - 2t^2 A(\pi_\varphi,t^2)}{2t^2 B(\pi_\varphi,t^2)} \, .
\eea
In any case, the variable $\pi_\varphi$ \eqref{eqforpiphi} decouples and satisfies the equation
\bea
\label{solpiphi}
\dot \pi_\varphi = 2t P(\pi_\varphi,t^2) \, ,
\eea
which can be solved, at least numerically. Later, we will propose examples where it can be solved analytically. 

We continue with the equation for $E^\varphi$ \eqref{Ephieq} which can now be solved according to
\bea
E^\varphi \; = \; \exp \left( 2 \int^t du \,  u \, \frac{S_{(1)}(u)}{S_{(2)}(u)} \right) \, ,
\eea
where $S_{(a)}(t)=S_{(a)}( \pi_\varphi(t),P(\pi_\varphi(t),t^2),t^2)$ and $\pi_\varphi(t)$ is  given by the solution of \eqref{solpiphi}. 
Of course, $E^\varphi$ is defined up to a constant which can be fixed by physical conditions.
Finally, the remaining variable $\pi_r$ is immediately obtained from \eqref{implicitP}.

\subsubsection{Summary of the results: metric in the black hole interior}
To conclude this section, we summarize the results. The resolution relies on the choice of the lapse function \eqref{lapsechoice} (which is equivalent to a change of coordinate) and on the (implicit) inversion of the Hamiltonian constraint which state the existence of a function $P$ such that
\bea
\frac{\pi_r}{E^\varphi} = P(\pi_\varphi, {E^r}) \, , \qquad \text{where} \quad 2 E^r S(\pi_\varphi, P(\pi_\varphi, {E^r}) , E^r) \; = \; 1 \, .
\eea
In general, $P$ is implicitly defined only and can be computed locally (at the vicinity of any points in the phase space when $S_{(2)}\neq 0$). In some cases, the function $P$ can be  found explicitly. Then,  the general solution is
\bea
\label{fullsol}
E^r=t^2 \, , \quad
\dot\pi_\varphi = 2t P(\pi_\varphi,t^2) \, , \quad
E^\varphi= \exp \left( 2 \int^t du \,  u \, \frac{S_{(1)}(u)}{S_{(2)}(u)} \right)  \, , \quad
\pi_r = E^\varphi P(\pi_\varphi,t^2) \, ,
\eea
which depends on two integration constants, the first one coming from the integration of $\pi_\varphi$, and the second one
coming in the integral defining $E^\varphi$. Notice that the constant $a$ which appears in the integration of $E^r$ \eqref{Ersolution}
has already been fixed to $a=0$. Finally, the metric inside the spherical black hole is given by
\bea
\label{lineel}
ds^2 = -\frac{1}{F(t)}dt^2 + G(t) \, dr^2 + t^2 (d\theta^2 + 
\sin^2\theta \, d\varphi^2) \, .
\eea
where the two functions $F$ and $G$ are defined by
\bea
\label{corrr}
F(t) \equiv \frac{S_{(2)}^2}{4} \, , \qquad
G(t) \equiv \frac{(E^\varphi)^2}{t^2}  \, .
\eea
Such a metric corresponds to a black hole interior if it admits at least one event horizon, which imposes conditions
on the functions $F$ and $G$.

\subsection{Conditions for describing a trapped interior region}

At this stage, we have obtained an exact expression of the most general homogeneous solution of the very general effective theory
introduced before. The components  of the effective metric are expressed in terms of the functions that govern the deformation of the
theory. Nonetheless, a priori, this solution could describe a cosmological background or a homogeneous black hole interior. 
Here, we are interested in the case of black holes interiors. 

In order for this solution to describe a well defined black hole interior, and therefore a trapped region, one has to impose some conditions.
\begin{itemize}
\item The geometry is bounded by an outer horizon located at $t_{\text{h}}$.
\item This outer horizon is null, and therefore it can be interpreted as a black hole horizon.
\item The geometry can be consistently extended through this horizon, in the outer-communication region corresponding to $t > t_{\text{h}}$.
\end{itemize}
In order to impose these three conditions, it is useful to introduce the Kodama vector defined for any spherically symmetric geometry as follows. Following \cite{Abreu:2010ru}, a general spherically symmetric spacetime can always be written as,
\be
ds^2 = \sigma_{ab} dx^a dx^b + R^2(t,r) d\Omega^2 ,
\ee 
where $\sigma_{ab}$ is the 2-dimensional metric on the base space with standard coordinates $(t,r)$,  the scalar function $R(t,r)$ is the physical radius, and $d\Omega^2$ is the metric on the normalized 2-sphere. With these notations, the Kodama vector is defined by
\be
k^{\alpha} : = \epsilon^{\alpha\beta}_{\perp} \nabla_{\alpha} R \;,
\ee
where $\epsilon^{\alpha\beta}_{\perp} = \epsilon^{\alpha\beta}/ \sqrt{|\sigma|}$ is the densitized 2-dimensional Levi-Civita tensor (see \cite{Abreu:2010ru, Kodama:1979vn} for details as well as \cite{Gegenberg:1994pv, Corichi:2015vsa, Grumiller:2007ju}). In time-dependent and inhomogeneous geometries, where there are no time-like Killing vector, this vector turns out to be especially useful to locate the horizons. For asymptotically flat spacetime, it coincides with the time-like Killing vector as spatial infinity. The crucial property of the Kodama vector is that it encodes the causal structure of the spacetime. In particular, assuming that there is a single horizon at $t_{\text{h}}$, surrounding a trapped region for $t < t_{\text{h}}$, the Kodama vector turns out to be time-like, null and space-like for $t > t_{\text{h}}$,  $t=t_{\text{h}}$ and $0< t < t_{\text{h}}$ respectively. As such the Kodama vector becomes space-like in a trapped region. 

Hence, we compute the Kodama vector in the homogeneous interior metric we found, and we obtain that
\be
k^{\alpha} \partial_{\alpha}= \frac{1}{\sqrt{|g_{tt} |g_{rr}}} \frac{\partial}{\partial r} =  - \sqrt{\frac{F(t)}{G(t)}} \frac{\partial}{\partial r} \qquad \Rightarrow \qquad k^{\alpha} k_{\alpha} = F(t) \, .
\ee
Imposing the first two conditions above imply that there exists $t_{\text{h}}$ such that
\be
\label{cond1}
F(t_{\text{h}}) =0 \, ,
\ee
which signals the presence of an outer horizon. 
Now imposing that there is a coordinate system in which one can extend the spacetime beyond the horizon is equivalent to demand that the determinant of the metric remains regular on the horizon and keep the same sign both inside and outside the horizon, which ensures that the metric remains Lorentzian in the whole spacetime. As the determinant is trivially given by
\be
\text{det}(g) = - \frac{G}{F}\;  t^4 \sin^2{\theta} \, ,
\ee
then, at the horizon where $F(t_{\text{h}}) =0$, one has to impose the condition
\be
\label{introC}
0 < \lim_{t\to t_{\text{h}}}  C(t) < + \infty \, , \qquad
C(t) \equiv \frac{G(t)}{F(t)} \, .
\ee
As a consequence, one obtains in addition that $g_{rr}(t_{\text{h}})=G(t_{\text{h}})=0$, which ensures that the outer horizon is indeed a black hole horizon.
Moreover, one has to impose that the deformation are such that the curvature invariants remain regular at the horizon.

{Once we have identified the generic conditions for the solution (\ref{lineel}) to describe a black hole interior, bounded by at least an outer horizon, it would be interesting  to derive the conditions for this geometry to be regular. However the general expression of the metric (\ref{lineel}) prevents us from providing sharp and useful conditions on this issue. We leave therefore this interesting direction for future works. }
\subsection{Examples}

To illustrate the previous results, we consider theories where $S$ is of the form \eqref{SAB}, i.e. it is an affine function of $\pi_r/E^\varphi$.  
In that case, the function $P$ entering in \eqref{fullsol} exists globally and can be computed explicitly. Furthermore, we assume that
\bea
A(\pi_\varphi,E^r) = -\frac{f_1(\pi_\varphi)}{2E^r} \, , \qquad
B(\pi_\varphi,E^r) = -2 f_2(\pi_\varphi) \, ,
\eea
where $f_1$ and $f_2$ are functions of $\pi_\varphi$ only. If one imposes the closeness of the constraints algebra, then $A$ and $B$
satisfy the condition \eqref{condAetB} which translates into $2f_2={df_1}/{d\pi_\varphi}$, and one recovers obviously the well-known anomaly-free
condition \cite{Tibrewala:2012xb,Bojowald:2015zha,BenAchour:2018khr}. To be general, we assume for the moment that $f_1$ and $f_2$ are independent.

Let us compute the building blocks to find the explicit form of the line element \eqref{lineel}. 
A direct calculation shows that
\bea
P=-\frac{1}{4E^r}\frac{1+f_1(\pi_\varphi)}{ f_2(\pi_\varphi)} \, , \quad
S_{(1)}= \frac{1}{2 E^r} \left( -f_1' +  \frac{1+f_1(\pi_\varphi)}{ f_2(\pi_\varphi)} f_2' \right)\, , \quad
S_{(2)}=-2f_2(\pi_\varphi) \, ,
\eea
where $f_a'$ ($a=1,2$) is the derivative of $f_a$ with respect to $\pi_\varphi$. Using $E^r=t^2$ and, after a short calculation, one shows that the equation for 
$\pi_\varphi(t)$ can be reformulated as
\bea
\frac{2 f_2(\pi_\varphi)}{1+f_1(\pi_\varphi)} \dot\pi_\varphi \; = \; - \frac{1}{t} \, .
\eea
On integrates easily this equation and shows that $\pi_\varphi$
is obtained by inverting the relation
\bea
\label{firsteq}
1+{f}(\pi_\varphi) = \frac{r_s}{t} \, , \qquad \text{with} \quad
{f}(\pi_\varphi) \equiv \exp \left(\int^{\pi_\varphi } \frac{2f_2(x)}{f_1(x)+1} dx \right)- 1\, ,
\eea
where $r_s$ is a constant of integration. Notice that the constant that comes from the integral in the r.h.s. of \eqref{firsteq} can be reabsorbed
into $r_s$.

The expression of $E^\varphi$ follows immediately and, after direct calculations, one obtains
\bea
E^\varphi(t) =  \frac{2b\, f_2(\pi_\varphi)}{1+ f_1(\pi_\varphi)} =  \frac{b}{1+{f}(\pi_\varphi)} f'(\pi_\varphi)= \frac{b t}{r_s} f'(\pi_\varphi) \, ,
\eea
where $\pi_\varphi(t)$ is given by \eqref{firsteq} and $b$ is a constant of integration. Without loss of generality, we can
fix $2b=r_s$ (as in \cite{BenAchour:2018khr}) by a redefinition of the constant in the integral defining $f$ \eqref{firsteq}.
As a conclusion, the line element is given by \eqref{lineel} with
\bea
F(t) = \left[ (f_2 \circ f^{-1})(\frac{r_s}{t}-1)\right]^2 \, , \qquad
G(t) = \frac{1}{4}\left[ (f' \circ f^{-1})(\frac{r_s}{t}-1)\ \right]^2 \, .
\eea
As $f'=2f_2/(f_1+1)$, we can simplify further the expression of $G(t)$ which can be written in the form $G(t)=C(t)F(t)$ as in \eqref{introC} with
\bea
\label{expofC}
C(t)=\frac{r_s^2}{t^2}{\left[ 1+ (f_1 \circ f^{-1})(\frac{r_s}{t}-1)\right]^{-2}} \, .
\eea

 Interestingly, as ${f}_1$ and $f_2$ are generically independent, $F$ and $G$  are themselves independent and 
 one obtains the most general ``static'' and spherically symmetry solution for the metric. 
 Such a geometry describes effectively a black hole interior if it admits an event horizon which imposes the conditions  we have
 described in the previous subsection. In particular, if we assume that the horizon is located at $t=r_s$ (as for the Schwarzchild solution),
 these conditions simplify and read
 \bea
( f_2 \circ f^{-1})(0) = 0 \, , \qquad
(f_1 \circ f^{-1})(0) \neq -1 \, .
 \eea
 In the case where the condition $2f_2=f_1'$ to have a closed constraints algebra is
satisfied, then ${f}+1=\lambda(f_1+1)$ where $\lambda$ is an integration constant.
A quick calculation shows that the expression \eqref{expofC} reduces to $C(t)=\lambda$ and 
one recovers the result of \cite{BenAchour:2018khr} 
\bea
F(t)=G(t) = \frac{1}{4}\left[ (f_1' \circ f_1^{-1})(\frac{r_s}{t}-1)\right]^2 \,,
\eea
where $\lambda$ has been fixed to $\lambda=1$.

\section{Discussion}

In this article, we have revisited and studied further the issue of deformed covariance in the so-called polymer models  of 
quantum spherically symmetric general relativity in vacuum. We have obtained several new results. 

First, in section \ref{sectionII}, we
introduced a large class of modified gravity theories which mimic, at the level of the full theory (hence beyond spherical symmetry), 
loop quantum gravity (holonomy like) deformations that one uses in standard symmetry reduced polymer models. In these theories,
the invariance under space-time diffeomorphisms is broken whereas 3 dimensional diffeomorphisms remain symmetries, as it is expected
from loop quantum gravity, and only the extrinsic curvature part of the Lagrangian is deformed compared to general relativity. Obviously,
these theories have a deformed Hamiltonian constraint. Hence, we ask the question whether it is possible that the constraints algebra remains closed even though it is deformed, and we show that this is impossible. In the full class of these theories of modified gravity, only general 
relativity leads to a closed constraints algebra, and such a no-go result is very similar to the  uniqueness Hojmann-Kuchar-Teitelboim theorem \cite{Hojman:1976vp}. As a consequence, it seems that the notion of deformed covariance found in  spherically symmetric model  and discussed in detail in \cite{Bojowald:2015zha}, could be a peculiar consequence of the spherical symmetry. 
Nevertheless, we must emphasize that the effective models we introduced are based on the implicit assumption that quantum gravity
effects can be described in terms of a local action whose dynamical variable is still a metric. Hence, it is in principle possible to evade the 
no-go result considering instead non-local actions or non-metric and more exotic variables. 

Second, we focus on spherically symmetric models. We consider the Hamiltonian theory whose phase space consists in two pairs 
of conjugate variables (schematically the two free functions in the most general spherically symmetric metric together with their conjugate momenta) which satisfy two constraints, the Hamiltonian and the vectorial constraint. We ask the question to which extend one can  deform these constraints, compared to general relativity, with the condition that their algebra remains closed. We show that there is no non-trivial
deformation of the vectorial constraint, which provides a very strong argument for keeping the vectorial constraint undeformed as it is usually
done in the context of polymer models. Then, we compute the most general deformation of the Hamiltonian constraint and we recover
known results in the literature. Our construction provides a large class of effective theories with a deformed covariance which encompass 
most of polymer models introduced in the literature. Notice that the model introduced recently in \cite{Ashtekar:2018lag, Bodendorfer:2019cyv} does not fall 
in this class mainly because the corresponding Hamiltonian constraint is, a priori, not a scalar with respect to the vectorial constraint. 

Finally, we compute the equations of motion and look for spherically symmetric static solutions. Remarkably, we found an explicit, exact 
and very general solution. In particular, we give the expressions of the components of the effective metric  in terms of the functions that govern the deformations of the theory. This exact solution, given by (\ref{lineel}) and (\ref{corrr}), can describe both a cosmological or a black hole interior solution. Assuming that the solution corresponds to the interior of a  black hole, we gave the general conditions (\ref{cond1}) and (\ref{introC}) in order for the effective geometry to come with a well-defined trapped region. 

This work opens many directions. First of all, one can extend our analysis to include different models as the ones introduced 
in  \cite{Ashtekar:2018lag, Bodendorfer:2019cyv}. One can try to find how to characterize more generally such models and to compute their general static 
solutions as we have done in the present work.  Moreover, we have now a framework to address the question of the stability of the solutions we found with respect to linear
(and spherical) perturbations. To our knowledge, stability properties of the perturbed polymer black holes have been investigated only in \cite{Boehmer:2008fz} and in \cite{Moulin:2018uap}. Let us finally stress that the present modified gravity model could also be used to derive polymer-like deformations of the spherically symmetric exterior geometry, and could be extended to describe  deformations of the axi-symmetric vacuum gravity phase space. This would allow us to discuss the effective dynamics of polymer deformations of the Kerr black hole (see \cite{Gambini:2018ucf} for current efforts in this direction). We leave these open directions for future works.

\bigskip
\textbf{Acknowledgements}\\
The work of J.BA was supported by Japan Society for the Promotion of Science Grants-in-Aid for Scientific Research No. 17H02890.
K.N acknowledges support from the CNRS project 80PRIME.


\end{document}